\documentclass[twocolumn,10pt]{article} % here we use the article class, rather than elsarticle

\usepackage[square,numbers,sort&compress,comma]{natbib}

\usepackage{amssymb}
\usepackage{caption}
\usepackage{graphicx}
\usepackage{latexsym}
\usepackage{times}
\usepackage[pagewise]{lineno}
\usepackage{tabularx} % Helpful for full text width tables.
\usepackage{graphicx} % Figures.
\usepackage{geometry}
\usepackage{tikz}
\usepackage[intlimits]{amsmath}

\usepackage{times}
\usepackage{soul} % Prevents underlined words/phrases from running into the margins.
\usepackage[square,numbers,sort&compress]{natbib} % References formatting.
\usepackage[labelfont=bf,font=normalsize]{caption}  % Bold captions.
\usepackage{subcaption}
\usepackage{epstopdf} % Uncomment if using PdfLaTeX command when building pdf.
\usepackage{hyperref} 
\usepackage{comment}
\usepackage{tabularx} % Helpful for full text width tables.
\usepackage{graphicx} % Figures.
\usepackage{geometry}
\usepackage{tikz}
\usepackage[intlimits]{amsmath}
\usepackage{amssymb} % Math symbols.
\usepackage{times}
\usepackage{soul} % Prevents underlined words/phrases from running into the margins.
\usepackage[square,numbers,sort&compress]{natbib} % References formatting.
\usepackage[labelfont=bf,font=normalsize]{caption}  % Bold captions.
\usepackage{subcaption}
\usepackage{epstopdf} % Uncomment if using PdfLaTeX command when building pdf.
\usepackage{hyperref} 
\usepackage{comment}
\hypersetup{
    colorlinks=true,
    citecolor = blue,
    linkcolor=blue,
    urlcolor = blue
}

\usepackage{amsmath}
\usepackage{amssymb}
\usepackage{caption}
\usepackage{graphicx}
\usepackage{latexsym}
\usepackage{times}
\usepackage[pagewise]{lineno}
\usepackage{tikz}

\usepackage{xcolor}
\usepackage{fmtcount}
\usepackage{marvosym}
\usepackage{marvosym}
\topmargin - 12pt % might need to be set to 0pt for some installations
\renewenvironment{abstract}%
              {% - begin definition
               \small% - select font
               {\bfseries \abstractname}% - select font
               \par% - end a paragraph (skip \parsep)
               \vspace{10pt}% - add vertical space
              }% - complete definition

\renewcommand\abstractname{Abstract}

\newcommand{\nomenclature}% - name of command
              [1]% - number of arguments
              {% - begin definition
               \bgroup% - begin a local group
               \flushleft% - turn on flushleft option
               \small\bf% - select font
               #1% - insert title text
               \par% - end a paragraph (skip \parsep)
               \egroup% - terminate local group
              }% - complete definition

\renewcommand{\section}% - name of command
              [1]% - number of arguments
              {% - begin definition
               \bgroup% - begin a local group
               \flushleft% - turn on flushleft option
               \small\bf% - select font
               \refstepcounter{section}% - increment counter
               \arabic{section}. #1% - insert title text
               \par% - end a paragraph (skip \parsep)
               \egroup% - terminate local group
              }% - complete definition

\renewcommand{\subsection}% - name of command
              [1]% - number of arguments
              {% - begin definition
               \bgroup% - begin a local group
               \flushleft% - turn on flushleft option
               \small\em% - select font
               \refstepcounter{subsection}% - increment counter
               \arabic{section}.% - insert title text
               \arabic{subsection}. #1% - insert title text
               \par% - end a paragraph (skip \parsep)
               \egroup% - terminate local group
              }% - complete definition

\renewcommand{\subsubsection}% - name of command
              [1]% - number of arguments
              {% - begin definition
               \bgroup% - begin a local group
               \flushleft% - turn on flushleft option
               \small\em% - select font
               \refstepcounter{subsubsection}% - increment counter
               \arabic{section}.% - insert title text
               \arabic{subsection}.% - insert title text
               \arabic{subsubsection}. #1% - insert title text
               \par% - end a paragraph (skip \parsep)
               \egroup% - terminate local group
              }% - complete definition

  \newcommand{\acknowledgement}% - name of command
              [1]% - number of arguments
              {% - begin definition
               \bgroup% - begin a local group
               \flushleft% - turn on flushleft option
               \small\bf% - select font
               #1% - insert title text
               \par% - end a paragraph (skip \parsep)
               \egroup% - terminate local group
              }% - complete definition

  \newcommand{\sectionbib}% - name of command
              [1]% - number of arguments
              {% - begin definition
               \bgroup% - begin a local group
               \flushleft% - turn on flushleft option
               \small\bf% - select font
               #1% - insert title text
               \par% - end a paragraph (skip \parsep)
               \egroup% - terminate local group
              }% - complete definition

\begin{document}

\title{\LARGE Effects of flame macrostructures on the combustion dynamics of novel counter-rotating radial swirl injector in a model can combustor}

\author{{\large SK Thirumalaikumaran$^{a}$, Balasundaram Mohan$^{a}$, Saptarshi Basu$^{a,b,*}$}\\[10pt]
        {\footnotesize \em $^a$Department of Mechanical Engineering, Indian Institute of Science Bangalore, India.}\\[-5pt]
        {\footnotesize \em $^b$Interdisciplinary Centre for Energy Research, Indian Institute of Science Bangalore, India.}\\[-5pt]
        {\footnotesize \em corresponding author: sbasu@iisc.ac.in}}

\date{}

% -------------------------------------------------------------------- %
% -------------------------------------------------------------------- %
% -------------------------------------------------------------------- %

\small
\baselineskip 10pt

% -------------------------------------------------------------------- %
% -------------------------------------------------------------------- %
% -------------------------------------------------------------------- %

\twocolumn[\begin{@twocolumnfalse}
\vspace{50pt}
\maketitle
\vspace{40pt}
\rule{\textwidth}{0.5pt}
\begin{abstract} % 100 to 300 words.
This study explores the flame macrostructures observed during self-sustained thermoacoustic oscillations in a model can combustor featuring a novel counter-rotating swirler. The swirler is designed to achieve high shear, distributing 60\% of the airflow through the primary passage and 40\% through the secondary passage, with radial fuel injection introduced via a central lance. To examine the influence of flame macrostructures on combustion instabilities, the flow expansion angles at the combustor's dump plane are systematically varied. In the present study, we consider three different flare angles comprising 90$^o$, 60$^o$, and 40$^o$ for Reynolds numbers and thermal powers ranging from $10500-16800$ and $10-16.2$ kW, respectively. Acoustic pressure, high-speed stereo PIV, and high-speed OH$^*$ chemiluminescence measurements are conducted to scrutinize flow and flame macrostructures during combustion instability. Large-amplitude acoustic oscillations are observed for a flare angle of \(40^\circ\), accompanied by shorter flames and wider central recirculation zones. In contrast, a flare angle of \(90^\circ\) results in low-amplitude acoustic oscillations characterized by longer flames and narrower central recirculation zones. The phase-averaged Rayleigh index is utilized to pinpoint regions that drive or dampen thermoacoustic instability, enabling the identification of potential strategies for its mitigation. Additionally, time-series analysis is employed to reveal the dominant acoustic modes and their interaction with heat release rate fluctuations.
\end{abstract}
\vspace{10pt}
\parbox{1.0\textwidth}{\footnotesize {\em Keywords:} Combustion instability; Partially premixed flames; High-shear injector; Counter-rotating radial swirler; Passive control}
\rule{\textwidth}{0.5pt}
\vspace{10pt}

\end{@twocolumnfalse}]

\clearpage

\twocolumn[\begin{@twocolumnfalse}

\centerline{\bf Information for Editors and Reviewers}

\vspace{20pt}

\vspace{20pt}

{\bf 1) Novelty and Significance Statement}
\vspace{10pt}

This study presents the combustion dynamics of a novel counter-rotating radial swirler, offering enhanced control over flame stabilization and flow recirculation, which is critical for stable operation. Through detailed analysis of combustion dynamics, including flame structure and acoustic coupling, the work provides a deeper understanding of instability mechanisms for different flame macrostructures. Notably, slender and elongated flame configurations demonstrated superior thermoacoustic stability compared to broader or more compact flame shapes. These findings suggest that implementing swirler clusters designed to promote slender flame stabilization could serve as an effective strategy for mitigating thermoacoustic instabilities in real-world combustion systems.

%The combustor operates under a partially premixed regime, closely replicating real-world gas turbine conditions, thereby improving the relevance and applicability of the findings. 

\vspace{20pt} 

{\bf 2) Author Contributions}
\vspace{10pt}

\begin{itemize}

  \item{SKT:} Conceptualization, Methodology, Investigation, Analysis, Writing – original draft, review \& editing. 

  \item{BM:} Conceptualization, Methodology, Investigation,  Analysis, Writing – review \& editing.

  \item{SB:} Funding acquisition, Supervision, Writing – review \& editing.

\end{itemize}

\vspace{20pt}

\end{@twocolumnfalse}] 

% -------------------------------------------------------------------- %
% -------------------------------------------------------------------- %
% -------------------------------------------------------------------- %

\clearpage

%\linenumbers

%\vspace{-5 mm}
\section{Introduction\label{sec:introduction}} \addvspace{10pt}

Swirlers are integral to gas turbine combustors, forming recirculation zones and reducing axial velocity to stabilize flames. These zones enhance mixing between hot reactive species and incoming mixtures \cite{SYRED1971}. In premixed flames, PVC enhances turbulence and compacts flames but minimally reduces thermoacoustic oscillations. In partially premixed flames, PVC reduces thermoacoustic instabilities by 80\%, improving mixing and mitigating equivalence ratio fluctuations \cite{Finn2019}. PVC-driven turbulence-chemistry interactions, including extinction, rapid flame recovery, and auto-ignition effects, are crucial near lean blowout (LBO) limits. Modifying flow or mixture fractions around critical regions extends LBO limits to leaner conditions \cite{STOHR2011, BOXX2010}.

Swirl motion influences mixing and flame structure, with axial injection causing turbulent interactions and intermittent fuel penetration, while radial injection eliminates this effect \cite{OLIVANI2007}. Flame stabilization and liftoff are influenced by inner swirl (S1), outer swirl (S2), and momentum ratio (J). Increasing S2 or reducing J promotes flame attachment, while higher S1 or J favors detachment. Adjusting the quarl angle or adding a diverging cup at the oxidizer outlet significantly alters flame topology, reducing liftoff height and widening recirculation zones. The quarl is identified as a crucial parameter for flame control \cite{Degeneve2019, Degeneve2018}. Straight quarls enhance inflow swirl and inner recirculation zones, while diverging quarls create distinct exit regions, reducing mixing. Increasing quarl angles widen central recirculation regions, reduce the flame length, and shift stagnation points upstream due to lower axial velocity and adverse pressure gradients \cite{ELBAZ2016}. Flare angles affect Central Turbulent Recirculation Zone (CTRZ) width and induce shear layer instability, generating turbulence \cite{Estefanos2015}.

In the past, investigations were focused on the effects of flare angle, primary and secondary air injection of dual-swirl injectors on the flow and flame dynamics. However, the novel high-shear counter-rotating swirl injector considered here with radial fuel injection exhibits better flow field characteristics as reported from our group in the context of non-reacting flows \cite{sonu2020}. In this article, we examine the reacting flow field characteristics, particularly focusing on the effects of flame macrostructures on the driving and damping of thermoacoustic instabilities. Note that the change in flare angle alters the flow field--for instance the forward stagnation point of CTRZ--and hence alters the relative position of the mean flame height from the injection location.

\begin{figure}[h!]
\centering
%\vspace{-5 mm}
\hspace{-10 mm}
\includegraphics[trim={10mm 55mm 0 35mm},width=0.54\textwidth]{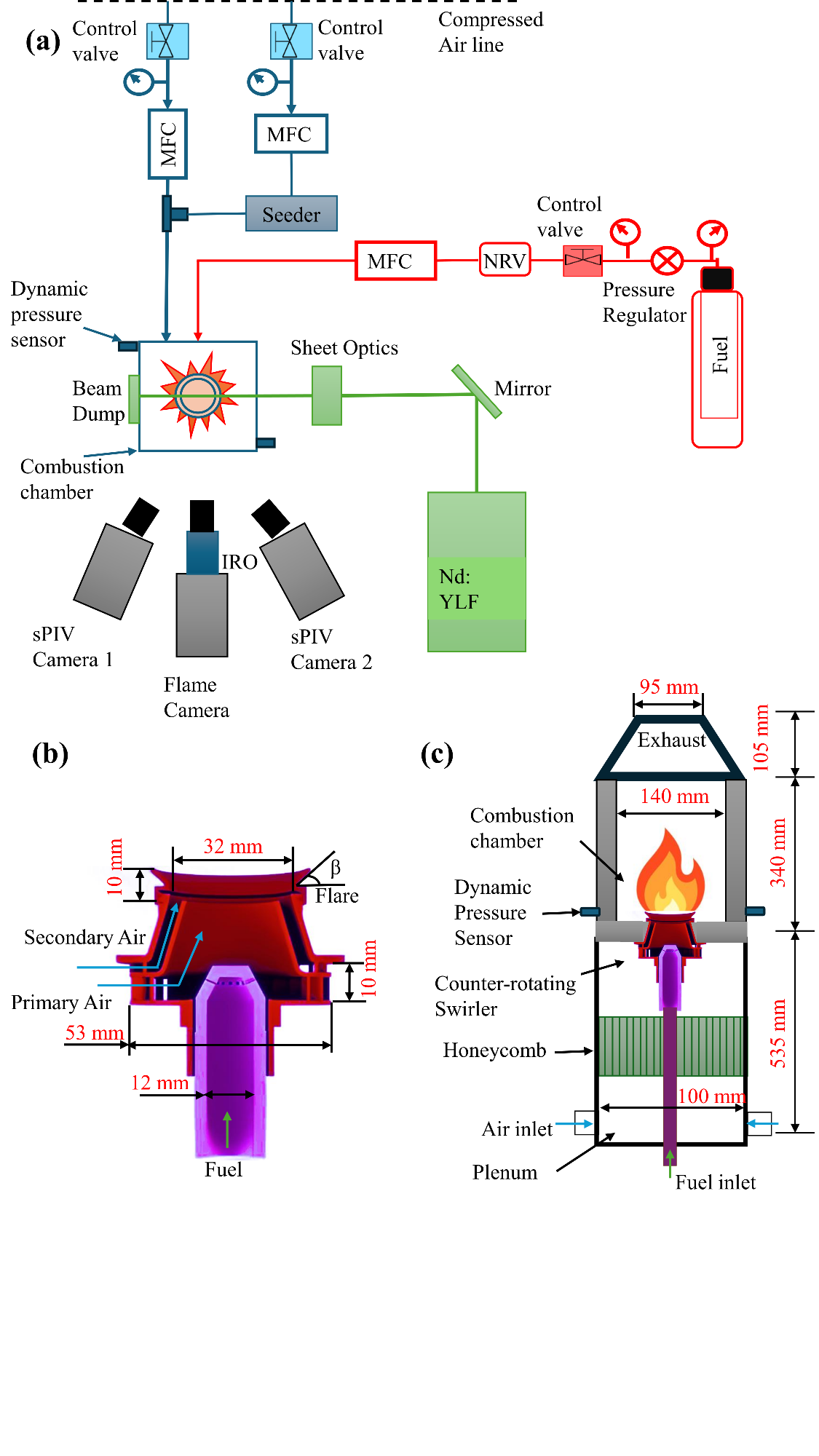}
\caption{\footnotesize (a) Schematic line diagram of the experimental setup, (b) Schematic of the swirler arrangement with the central lance, and (c) Schematic of the Combustor arrangement.}
\label{fig_experimental_setup}
\end{figure}

This article begins by outlining the experimental setup and diagnostic tools. Combustion dynamics are analyzed via acoustic pressure time-series data to determine dominant acoustic mode characteristics. Mean and phase-averaged flow/flame dynamics are then detailed, followed by a conclusion summarizing key findings.

\vspace{-1 mm}
\section{Combustor setup, data acquisition and processing\label{sec:sections}} \addvspace{10pt}

\begin{figure*}[h!]
\centering
    \vspace{-0.5 in}
    \includegraphics[width=0.6\textwidth]{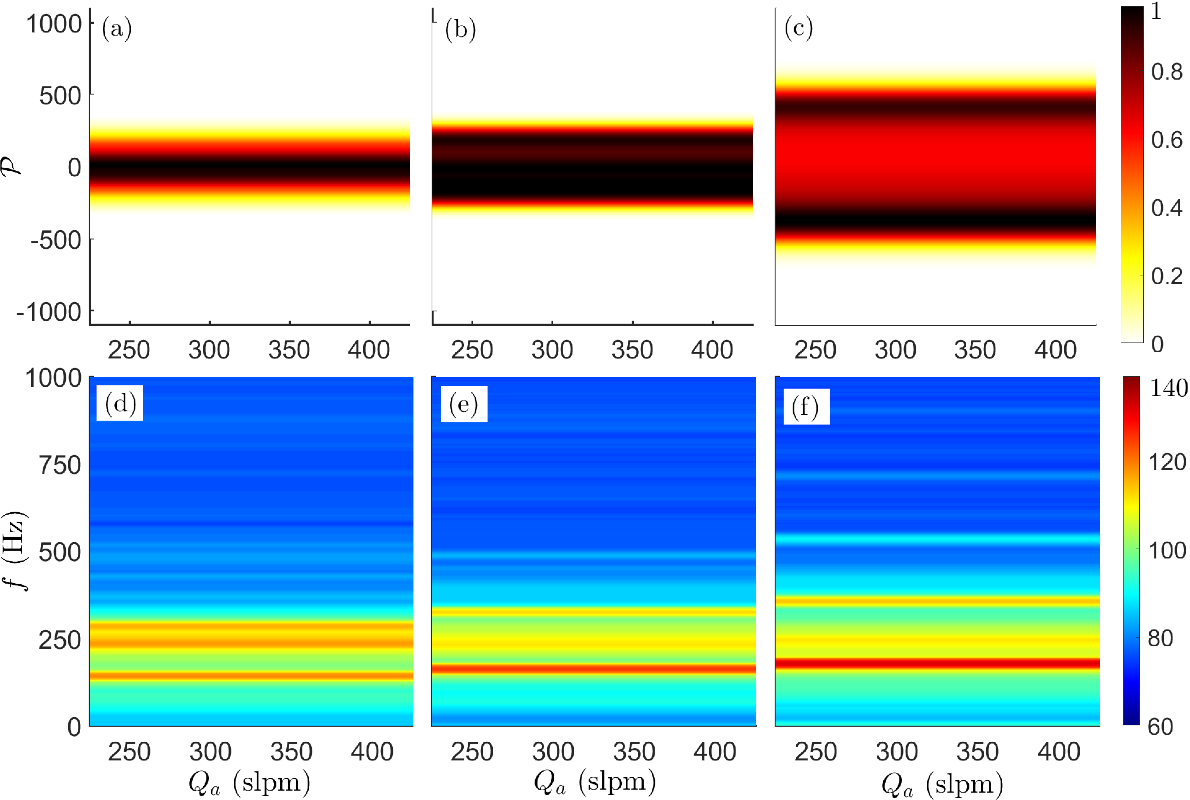}
    \vspace{1 pt}
    \caption{Probability density function (top row) and power spectral density in dB (bottom row). For flare angles (Left) $\beta=90^o$, (middle) $\beta=60^o$, and (right) $\beta=40^o$, respectively.}
	\label{fig_Pr_PDF_PSD}
    \vspace{-5 mm}
\end{figure*}

Experimental studies are conducted in a model can combustor comprising of a plenum, swirler, combustion chamber, and exhaust nozzle. Compressed air from a high pressure storage tank was regulated using an Alicat mass flow controller (MFC) (range: 0 to 500 slpm) and directed into a 0.1 m air plenum with a honeycomb structure for uniform flow. The air then enters counter-rotating swirlers with a 6:4 split ratio, where 60\% passes through the primary vanes and 40\% through the secondary vanes. Methane was used as fuel, injected through a centrally placed tube and regulated by another Alicat MFC (range: 0–50 slpm). Air and methane are mixed in primary and secondary regions before exiting through a flare (a diverging section) to achieve partially premixed combustion. The experimental setup with associated flow lines and instrumentation is depicted in figure \ref{fig_experimental_setup} (a). The model can combustor arrangement and the injector cut section are shown in figure \ref{fig_experimental_setup} (b) and (c).  The combustion chamber (0.14 m × 0.14 m × 0.34 m) featured fused silica windows of UV grade for flame and flow visualisation. Product gases are exited through a 0.095 m × 0.095 m converging nozzle. Experiments were conducted under atmospheric conditions for the Reynolds number of $10500-16800$ with a fixed equivalence ratio of 0.7. The Reynolds number is calculated as follows,
\vspace{-1 mm}
\begin{equation}
Re = \frac{\rho V D_s}{\mu}
\end{equation}
where $D_s$ is the swirler outlet diameter. The exit velocity $V$ of the swirler is calculated from the inlet mass flow rate of air. The swirler comprises a primary clockwise swirler and a secondary counter-clockwise swirler viewed from the dump plane, having geometric swirl numbers of 1.5 and 0.78, respectively. It accommodates a fuel injection nozzle with 12 radial jets (0.6 mm diameter) positioned concentrically, oriented 55$^\circ$ to the nozzle axis.

A high-speed stereo Particle Image Velocimetry (sPIV) system captures the flow field within the combustion chamber. Two high-speed SA-5 cameras (7000 frames per second, 1024 x 1024 pixels) from LaVision are used for imaging. The cameras operate at a frame rate and resolution of 3500 Hz and 1024 x 1024 pixels, respectively. The cameras have AT-X M100 Tokino f/2.8 camera lens, Scheimpflug adapters, and 527nm bandpass filters. The flow is seeded with micron-sized alumina particles (approximately 1 µm). A high-speed dual head Nd: YLF laser (DM30 series with 30 mJ per pulse energy, 527 nm wavelength, 10kHz) illuminates these seeder particles. The cylindrical laser beam is transformed into an approximately 1 mm thin laser sheet using a combination of telescopic lenses and a cylindrical lens within the sheet optics. The flow field data are post-processed using commercial Davis 8.4 software to obtain a velocity vector field. The processing method employs a multi-pass approach with an initial interrogation window of 64 x 64 pixels with 50$\%$ overlap, and a final interrogation window of 32 x 32 pixels with 75$\%$ overlap is utilized.

A LaVision SA-5 high-speed camera (1024 × 1024 pixels, 7000 fps) with High-Speed Intensified Relay Optics (HS-IRO) captures OH$^*$  chemiluminescence from the flame using a UV-sensitive lens (Cerco, 105 mm, f/2.8) and a $310\pm10$ nm bandpass filter. The HS-IRO is triggered via a LaVision Programmable Timing Unit (PTU-x) and recorded at 3500 Hz over 1 second, generating 3500 OH$^*$ images. A PCB 113B28 high-frequency pressure transducer (sensitivity of 100 mV/Pa) is mounted 10 mm from the dump plane for acoustic pressure measurements. Data is acquired at 35 kHz through an NI 9205 module in an NI cDAQ 9179 chassis. Further data processing is performed using MATLAB software.

\section{Results and discussion}

In the following subsections, we characterize the combustion dynamics of higher shear swirl injector and elucidate the underlying combustion dynamics.% using acoustic pressure and high-speed flame images primarily.

\begin{figure*}[h!]
	\centering
    \vspace{-0.5 in}
	\includegraphics[trim={0 1cm 14cm 0},width=0.85\textwidth]{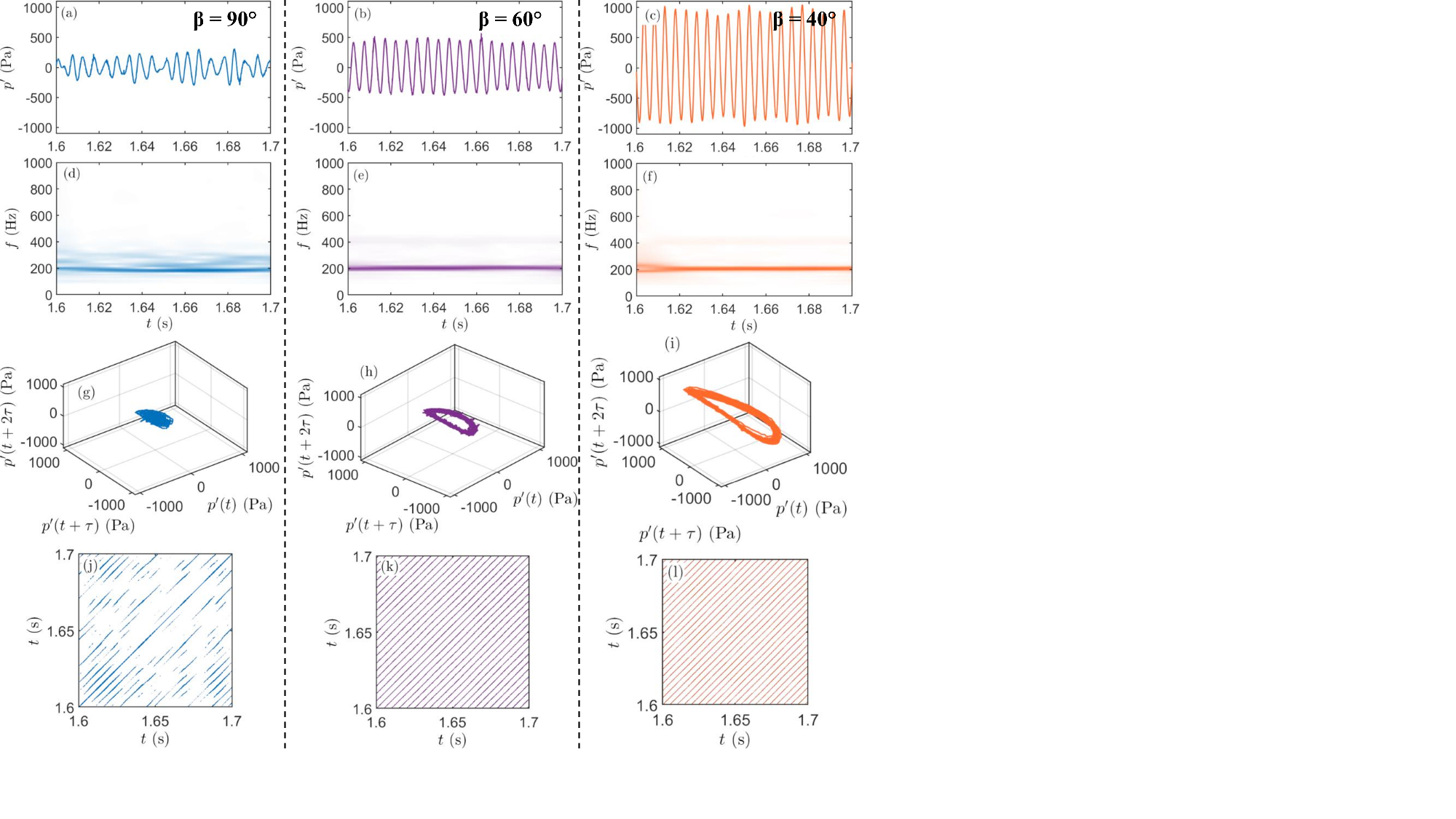}
	\caption{Left, middle and right columns of figures correspond to flare angles of 90$^o$, 60$^o$ and 40$^o$, respectively. (a-c) acoustic pressure, (d-f) wavelet scalogram, (g-i) phase portrait, and (j-l) recurrence plot, respectively.}
	\label{fig_TS_PSD_PP_RP}
    \vspace{-5 mm}
\end{figure*}

\subsection{Preliminary characterization \label{sec_char}}
The acquired acoustic pressure ($p^\prime$) and instantaneous OH$^*$ chemiluminescence ($\tilde{I}$) signals are further processed for characterization in this subsection. The probability density function ($\mathcal{P}$) of acoustic pressure field is used to obtain the probability of occurrence of different events ranging from low amplitude to large amplitude limit cycle oscillations. It is estimated using a non-parametric kernel density estimator. Further, the Gaussian kernel and optimal bandwidth associated with a method of normal reference rule are utilized in estimating the $\mathcal{P}$. The power spectral density of acoustic pressure field ($p^\prime$) and the temporal evolution of the frequency are analyzed using a Hann window with segment averaging and continuous wavelet transform, respectively. Bump wavelet is used to estimate the continuous wavelet transform.

%In figure \ref{fig_Pr_PDF_PSD}, the top and bottom rows show the probability density function (PDF) and power spectral density of the acoustic pressure acquired in the combustion chamber. The columns from left to right correspond to the flare angles ($\beta$) of $90^o$, $60^o$, and $40^o$, respectively. 
Figure \ref{fig_Pr_PDF_PSD}(a) shows the Gaussian distribution that indicates stable operation for all flow rates considered in $\beta=90^o$. Reducing the flare angle further leads to the spread in the Gaussian distribution around low amplitude fluctuations for $\beta=60^o$ as shown in panel (b). Subsequently, decreasing the flare angle to 40$^o$ shows large amplitude limit cycle oscillations with a bimodal distribution. Figure \ref{fig_Pr_PDF_PSD}(d-f) shows as the amplitude level observed increases to the larger values in figure \ref{fig_Pr_PDF_PSD}(a-c), the associated dominant frequency shifts slightly to higher values. In addition, stronger harmonics of the fundamental mode start to appear. This is further discussed in the ensuing paragraphs.

Phase space \cite{juniper2018sensitivity} is reconstructed from the acoustic pressure using time delay $(\tau)$ and embedding dimension ($d$). The $\tau$ is determined from the first local minimum of the average mutual information, while the embedding dimension is found using Cao's method. It is an optimized version of the false nearest neighbors technique. The $j-th$ vector reconstructed from the acoustic pressure can be expressed as follows.
\begin{equation}
y_j=(p^\prime(j),p^\prime(j+\tau),...,p^\prime(j+(d-1)\tau)).
\end{equation}
The threshold-dependent recurrence matrix is used to graph the Recurrence plot (RP) in this article. It is expressed as,
\begin{equation}
R_{m,n}=\Theta(\epsilon-||y_m-y_n||) ~~~~\&~~~~ m,n=1...N-(d-1)\tau
\end{equation}
where $R_{m,n},~\Theta,~\epsilon,~||.||$ and $N$ denotes the recurrence matrix, Heaviside function, threshold value, $l_2-$ norm and the total number of data points, respectively. $\epsilon$ is chosen as $20\%$ of the maximum diameter of the attractor; obtained from the phase space reconstruction. The recurrence matrix consists of binary values 0 and 1, represented by the white and colored points, respectively. In RP, homogeneous points, diagonal lines, and cell-like structures denote stochastic, periodic, and bursting fluctuations, respectively \cite{marwan2007recurrence}. The nonlinear characterization of the portion of time series for the airflow rate of $Q_a=350$ slpm is shown in figure \ref{fig_TS_PSD_PP_RP} for all three flare angles. The left column of figures correspond to $\beta=90^o$, shows low amplitude pressure oscillations. The wavelet scalogram shown in panel (d) shows a dominant peak at around 200 Hz with distributed frequency around it. The phase portrait in panel (g) shows a distributed structure attributed to the noise. The recurrence plot in panel (j) shows the discontinuous lines along the diagonal, illustrating instability amplitude is in the range of noise. On the other hand, the flare angle $\beta=40^o$ shown in the right column of figure \ref{fig_TS_PSD_PP_RP} exhibits large amplitude oscillations. The wavelet scalogram in panel (f) exhibits a sharp discrete peak around 200 Hz with a harmonic near 400 Hz. The phase portrait in panel (i) is a ring-like structure attributed to the limit cycle oscillation with the cycle-to-cycle variation attributed to the combustion noise. The recurrence plot in panel (l) shows clear diagonal lines related to the periodicity of the limit cycles observed in panel (c). The middle row of figure \ref{fig_TS_PSD_PP_RP} corresponding to flare angle $\beta=60^o$ manifests similar features observed for $\beta=40^o$.

\begin{figure*}[h!]
	\centering
    \vspace{-0.5 in}
    \includegraphics[width=0.65\textwidth]{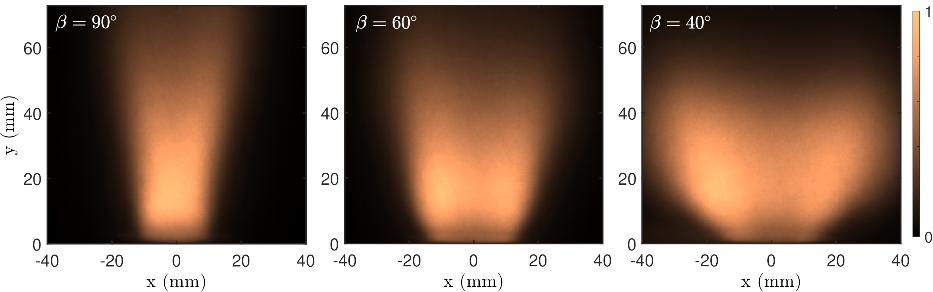}
    \includegraphics[width=0.65\textwidth]{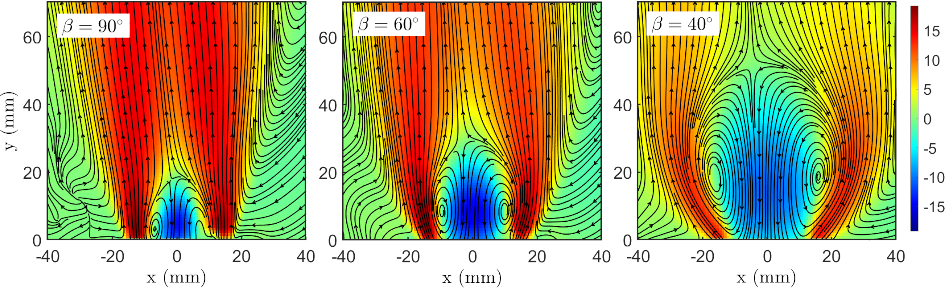}
	\caption{(Top row) Mean flame images normalized with their respective maximum values. (Bottom row) Axial velocity ($\bar{v}_y$ in m/s) with streamlines overlaid for clarity depicting recirculation zones, shear layers, and jet. }
	\label{fig_Flm_Vel}
    \vspace{-5 mm}
\end{figure*}

The top row of figure \ref{fig_Flm_Vel} depicts the mean flame image for three different flare angles. For $\beta=90^o$, the flame appears slender and elongated along the combustion chamber, whereas for $\beta=40^o$, it spreads out laterally and shrinks longitudinally. Intermediate lateral spread and longitudinal elongation were observed for $\beta=60^o$. The horizontally integrated mean heat release shows the maximum heat release occurs close to the dump plane for $\beta=90^o$ and axially spread (similar to long tail distribution). Relatively maximum mean heat release away from the dump plane occurs for $\beta=40^o$ with subsequent rapid drop along the axial direction. Further, the $\beta=60^o$ also exhibits a nearly similar mean heat release pattern to that of $\beta=40^o$. Therefore, the variation of flare angle alters the mean flame heat release distribution from axially concentrated to radially distributed. The fluctuations overlaid on this distribution can influence combustion dynamics significantly depending on the injector manifold impedance or acoustic boundary condition upstream of the dump plane.

The total velocity field, overlaid with streamlines, is presented in the bottom row of the figure \ref{fig_Flm_Vel}. At $\beta$ = 90$^\circ$, the axial velocity is maximized, and radial velocity is minimized due to the presence of a narrow central recirculation zone, as depicted in figure \ref{fig_Flm_Vel}. In contrast, $\beta$ = 40$^\circ$ exhibits a broader central recirculation zone, resulting in higher radial velocity and reduced axial velocity. The $\beta$ = 60$^\circ$ case demonstrates intermediate behavior, with moderate radial and axial velocities.
The dimensions of the central recirculation zone vary significantly with different flare angles. For $\beta$ = 40$^\circ$, the axial extent of the recirculation zone is approximately 60 mm, whereas for $\beta$ = 90$^\circ$, it is notably smaller, around 19 mm. This variation highlights the influence of $\beta$ on the velocity distribution and the size of the central recirculation zone. The broader central recirculation zone at $\beta$ = 40$^\circ$ impacts the flame dynamics, causing it to concentrate near the combustion chamber dump plane.

\subsection{Flame response and Rayleigh index}

\begin{figure*}[h!]
	\centering
    \vspace{-0.5 in}
	\includegraphics[width=0.8\textwidth]{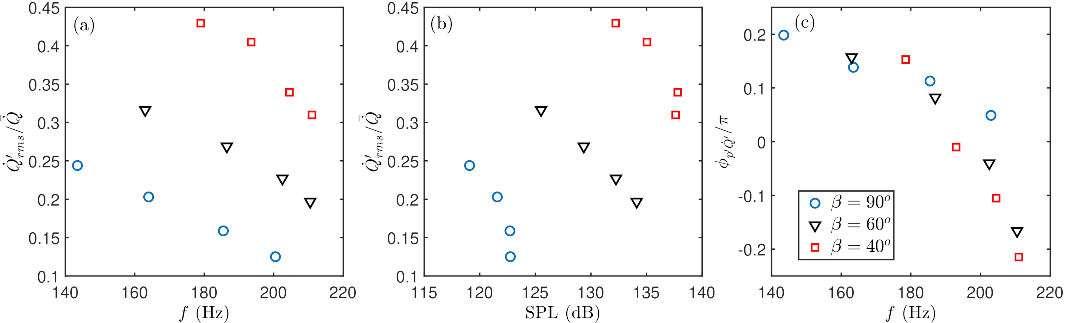}
	\caption{Normalized heat release rate response at the (a) dominant frequency ($f$) and associated (b) sound pressure level (SPL). (c) Normalized phase between acoustic pressure and heat release rate fluctuation at the dominant frequency falling between $\pm0.5$ denotes thermoacoustic driving. In panels (a-c), the Reynolds number and thermal power increase along the positive x-axis.}
	\label{fig_HRR_Resp}
    \vspace{-5 mm}
\end{figure*}       

Figure \ref{fig_HRR_Resp}(a) shows the Root Mean Square (RMS) of heat release rate fluctuations normalized with their respective mean for different flow rates versus the dominant acoustic mode frequency. It is interesting to note that increasing heat release rate fluctuations occur as the flare angles are decreased across all flow rates explored in this study. However, increasing the flow rates tends to narrow the difference in heat release rate fluctuations between two different flare angles. On the other hand, increasing thermal power and Reynolds number for a given flare angle leads to reduced heat release rate fluctuations. Furthermore, a relative increase in instability frequency leads to a relative reduction in heat release rate fluctuations for a given flare angle. This could be attributed to the low pass behavior of flame frequency response, where the flame heat release rate tends to decrease as the frequency increases beyond the cut-off frequency. In addition, fixed thermal power and Inlet Reynolds number with varying flare angles exhibit an increase in the instability frequency for the low flow rates, whereas for high flow rates, a non-monotonic dependence is observed. This increase in frequency with flow rate could be attributed to the reduced convective time of flow and equivalence ratio fluctuations to reach average flame height from the fuel injection location. Note that a negligible change in mean flame heat release rate shift along the axial direction is observed for a given flare angle for the flow rates considered in this study. In addition, as we have shown in figure \ref{fig_Flm_Vel}(top row), the mean flame length for a reduced flare angle for a fixed thermal power and Reynolds number reduces, leading to lower convective delay, causing higher frequency. Therefore, the difference in frequency across flare angles are attributed to the convective time, which in turn related to the changes in flame macro structures. A model to reproduce this frequency variation with flame macrostructure will be reported elsewhere.

In figure \ref{fig_HRR_Resp}(b), the RMS of normalized heat release rate fluctuations with associated Sound Pressure Level (SPL) corresponding to the dominant instability frequency are shown. Again, it is clear that across the flare angles for fixed thermal power and Reynolds number, increasing instability amplitude leads to an increase in flame frequency responses. Note considering all the flow rates explored in this study, the maximum mean intensity variation of 30\% was observed. The reduction in normalized heat release rate fluctuations could be attributed to the increased mean intensity values associated with the operating condition. However, during large amplitude acoustic oscillations, higher harmonics with considerable amplitude values are observed in heat release rate for $\beta=40^o$. The energy leakage from instability mode to higher harmonics through nonlinear heat release rate and acoustic pressure coupling could be attributed to non-monotonic variation in heat release rate at the dominant instability frequency (figure not shown).

\begin{figure*}[h!]
	\centering
    \vspace{-0.5 in}
	\includegraphics[width=0.83\textwidth]{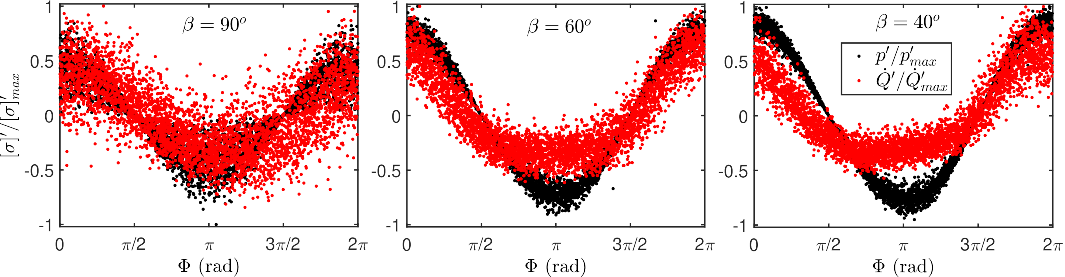}
	\caption{Acoustic pressure and HRR fluctuations normalized with their respective maximum values versus acoustic pressure phase angle ($\Phi$) for (Left) $\beta=90^o$, (middle) $\beta=60^o$, and (right) $\beta=40^o$, respectively denoting their relative phase and amplitude.}
	\label{fig_Pr_HRR_PA}
    \vspace{-5 mm}
\end{figure*}

The relative phase between the acoustic pressure and heat release rate fluctuations determines the thermoacoustic driving and damping. If the phase is between $\pm \pi/2$ thermoacoustic driving is encouraged otherwise damped. Note, however, for thermoacoustic instability to occur, the driving must overcome damping from thermal and viscous dissipation from the walls and acoustic losses at the boundary. The interpretations are made here assuming constant fixed damping in our combustor. The normalized phase (with $\pi$) estimated at the dominant instability frequencies for different flare angles are shown in figure \ref{fig_HRR_Resp}(c). It is interesting to point out that all the phase angles fall within $\pm\pi/2$ denoting the possibility of driving. However, we have shown in figure \ref{fig_TS_PSD_PP_RP}(left column) for $\beta=90^o$ and $Q_a=350$ slpm that the instability amplitude is in the order of background noise level. This in particular was clear from the phase portrait and recurrence plot as discussed. To further understand the cycle-to-cycle variation of the acoustic pressure relative to heat release rate fluctuations, they are related through acoustic pressure phase angle ($\Phi$) for the entire signal acquired. This is shown in figure \ref{fig_Pr_HRR_PA} for all flare angles. In figure \ref{fig_Pr_HRR_PA}, the acoustic pressure and heat release rate fluctuations are normalized with their respective maximum value for clarity. For $\beta=90^o$, as we have seen the background noise level is in the same order of acoustic pressure amplitude. Hence a significant cycle-to-cycle variation in acoustic pressure and heat release rate fluctuations are observed. However, the heat release rate appears to lag the acoustic pressure slightly and is nearly in phase. The $\beta=60^o$ case exhibits relatively better in-phase behavior than the other two flare angles considered. Heat release rate fluctuations lead the acoustic pressure in $\beta=40^o$ case. Also, the cycle-to-cycle variation of signal amplitudes was observed for large amplitudes as well in both signals. To get further insight into the driving and damping caused by the flame microstructures, the Rayleigh index is estimated as discussed below.

\begin{figure}[h!]
	\centering
    \hspace{-3 mm}
	\includegraphics[width=0.488\textwidth]{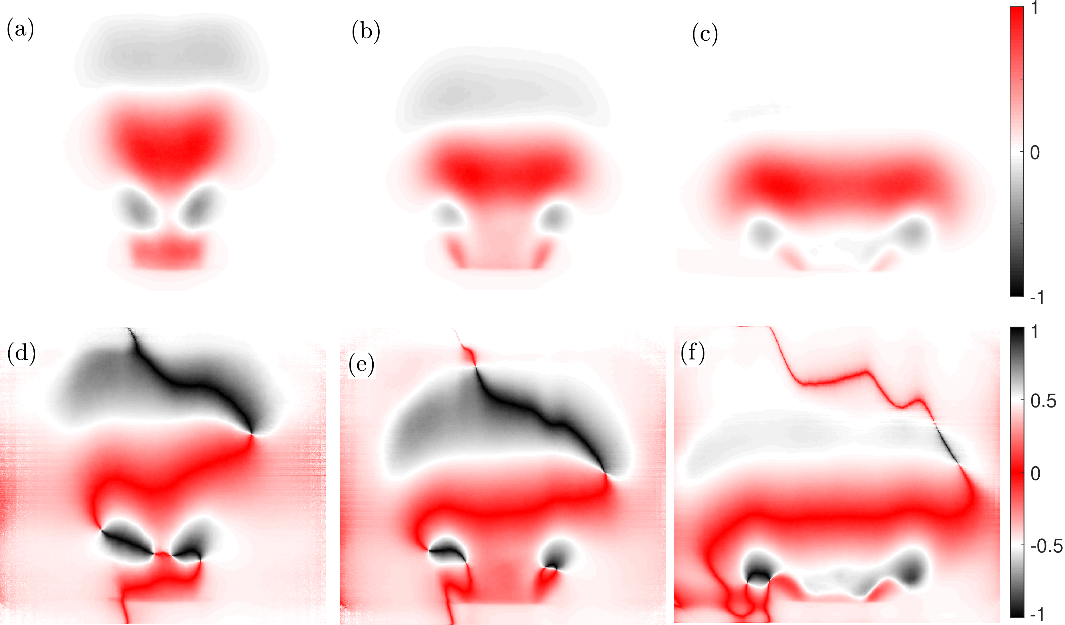}
    \vspace{2 pt}
	\caption{(Top row) Time averaged Rayleigh index (using Eq. \ref{eq_RI}) normalized with their respective maximum. (Bottom row) Phase angle associated with the top row of figures illustrating locations of driving ($<\pm 0.5$) and damping ($>\pm 0.5$), respectively. Flare angles (a,d) $\beta=90^o$, (b,e) $\beta=60^o$, and (c,f) $\beta=40^o$, respectively.}
	\label{fig_Avg_RI}
    \vspace{-5 mm}
\end{figure} 

The Rayleigh index ($\mathcal{R}$) expressed below sheds light on the spatial locations of driving and damping and their relative contribution to overall thermoacoustics caused by the flame macrostructures.
\begin{equation}
    \mathcal{R}(x,y,t)= \frac{\gamma -1}{\gamma \bar{p}} p^\prime(x,t) \dot{Q}^\prime(x,y,t) \label{eq_RI}
\end{equation}
In the above equation, $\gamma$ and $\bar{p}$ denote the ratio of specific heat capacity and mean combustion chamber pressure, respectively. Further, the spatial coordinates and time are denoted by $x,~y$, and $t$, respectively. The time averaged Rayleigh index and associated phase angle for all the flare angles considered in the study are shown in figure \ref{fig_Avg_RI}. Note, in figure \ref{fig_Avg_RI}(top row), the individual panels are normalized with their respective maximum values. It is clear from the acoustic pressure characterization shown in figure \ref{fig_Pr_PDF_PSD} that the relative acoustic pressure amplitude increases with the flare angle and further as discussed above the relative phase between pressure and heat release rate favors thermoacoustic driving. Therefore, with decreasing flare angles, the relative magnitudes of the total Rayleigh index increase significantly from $\beta=90^o$ to $40^o$. However, this cannot be interpreted from figure \ref{fig_Avg_RI} as the present study focuses on spatial locations of local driving and damping. Furthermore, it is interesting to note that the combustor supports pressure anti-node at the dump plane and pressure node at the combustor exit similar to that reconstructed in \cite{vishwakarma2022experimental}. Therefore, as shown in figure \ref{fig_Flm_Vel} the flame with maximum heat release rate close to the dump plane i.e., $\beta=40^o$ exhibits relatively large time averaged Rayleigh index and other flare angles exhibit relatively low value. Further, in contrast to the maximum mean heat release spatial location (refer to figure \ref{fig_Flm_Vel}), the maximum in Rayleigh index occurs downstream of mean heat release in all the flare angles (flame structures) explored in this study. The appearance of the Rayleigh index maximum in figure \ref{fig_Avg_RI} is reversed with respect to the maximum heat release considering the flare angles. Further, all the mean Rayleigh index profiles look the same. Corresponding phase angles are shown in figure \ref{fig_Avg_RI}(bottom row), the red shaded region denotes the driving ($\phi_{p^\prime \dot{Q}^\prime}<\pm \pi/2$) and the black shaded region represents damping ($\phi_{p^\prime \dot{Q}^\prime}>\pm \pi/2$), respectively. To understand further, the phase average Rayleigh index is discussed in the ensuing paragraphs for $\beta=90^o$ and $\beta=40^o$ in detail.

\begin{figure*}[h!]
	\centering
    \vspace{-0.5 in}
	\includegraphics[width=0.6\textwidth]{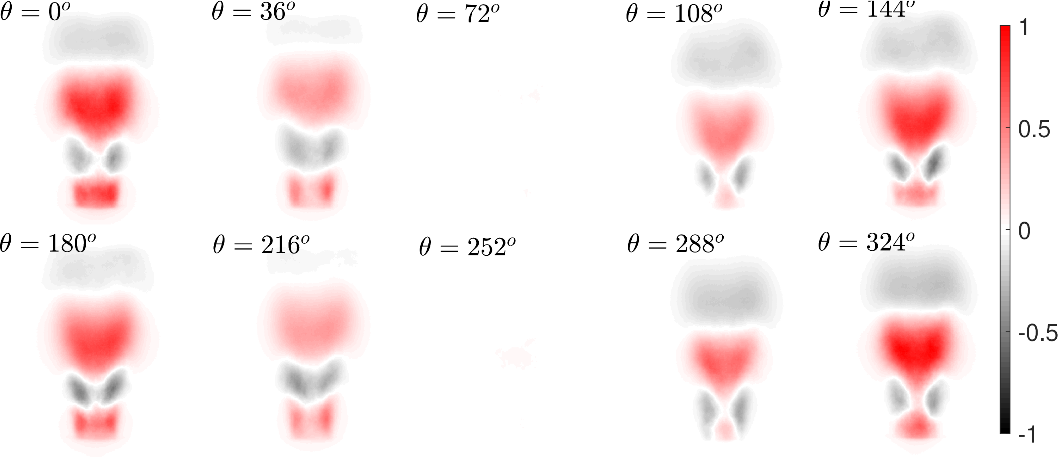}
    \vspace{5 pt}
	\caption{Phase averaged normalized Rayleigh index obtained using Eq. \ref{Eq_PA_RI} denotes the various spatial locations of damping and driving for $\beta=90^o$.}
	\label{fig_RI_FA90}
    \vspace{-2 mm}
\end{figure*}    

\begin{figure*}[h!]
	\centering
	\includegraphics[width=0.6\textwidth]{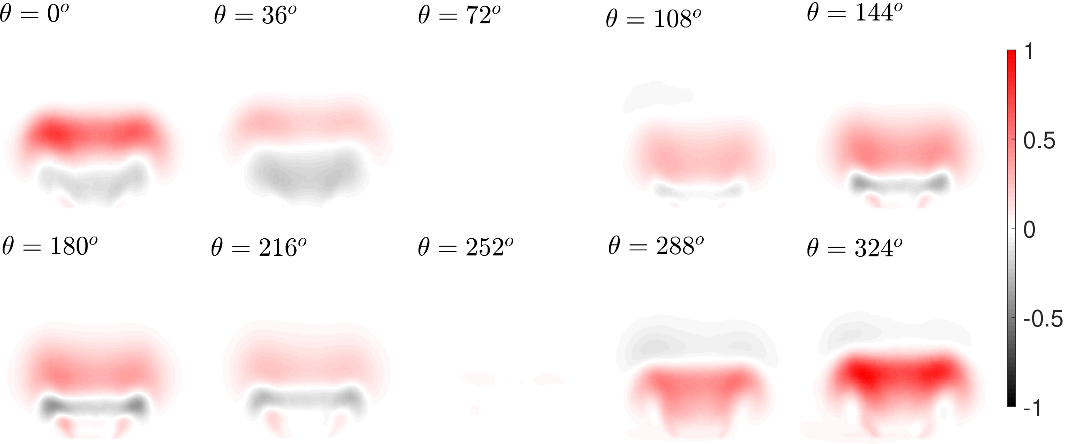}
    \vspace{5 pt}
	\caption{Phase averaged normalized Rayleigh index obtained using Eq. \ref{Eq_PA_RI} denotes the various spatial locations of damping and driving for $\beta=40^o$.}
	\label{fig_RI_FA40}
    \vspace{-5 mm}
\end{figure*}

The phase averaged Rayleigh index ($\mathcal{R}$) images are obtained in the present study to elucidate the driving and damping of thermoacoustic oscillation over a dominant instability cycle \cite{Bala_phd_thesis}. The $j-th$ phase averaged Rayleigh index is expressed as follows.
\begin{equation}
 \overline{\mathcal{R}}_j(x,y)=\langle \widetilde{\mathcal{R}}(x,y,t)\rangle _{(j-1)\pi/5 <\Phi(t) \le j\pi/5 } \label{Eq_PA_RI}
\end{equation}
where $j=1,...,10$. The arithmetic mean (denoted by $<[~]>$)  in the above equation denotes conditional averaging over a phase angle of 36$^o$. Note that the phase $\Phi(t)$ is estimated from the analytical signal of acoustic pressure fluctuations. The calculated phase averaged Rayleigh indexes are shown in figures \ref{fig_RI_FA90} and \ref{fig_RI_FA40} for flare angles 90$^o$ and 40$^o$, respectively. 

Figure \ref{fig_RI_FA90} shows the phase averaged Rayleigh index for 10 phase angles. The phase angles $72^o$ and $252^o$ fall close to the acoustic pressure node. Note, in figures \ref{fig_RI_FA90} and \ref{fig_RI_FA40}, the minimum phase angle associated with the bins is marked in the figure as a representative phase angle. In figure \ref{fig_RI_FA90}, the driving occurs very close to and far away from the dump plane separated by a damping. This damping location coincides with the location of the maximum mean heat release rate shown in figure \ref{fig_Flm_Vel}. Furthermore, comparing with the PIV mean flow field in figure \ref{fig_Flm_Vel}, the phase averaged Rayleigh index shows the inner recirculation zone drives the thermoacoustic instability irrespective of the phase angles. Furthermore, the flame stabilized in the jet emanating at $\theta=288^o$ contributes to the damping of thermoacoustic instability. As the phase evolves, this forms the necking and subsequently decoupling the driving occurring in the inner recirculation zone and its wake. It is clear from figure \ref{fig_RI_FA90} that the wake portion of the flame contributes predominantly to the driving of the thermoacoustic instability at $\beta=90^o$.

Comparing figure \ref{fig_Avg_RI} corresponding to $\beta=40^o$ and associated phase averaged Rayleigh index in figure \ref{fig_RI_FA40} shows the initial portion of jet and later portion of inner shear layer contributes predominantly to the driving of the thermoacoustic instability. Further, a thin band of acoustic damping in the tip of the jet observed between $\theta=108^o$ to 216$^o$ contributes predominantly to disrupting the coupling. Further more the majority of the driving occurs away from the dump plane. The driving occurring close to the dump plane at the later phase angle is nullified by the damping occurring at the initial phases. Therefore, $\beta=40^o$ the majority of the driving occurs in the region of the inner shear layer, and minimal contribution from the jet is observed. There is no contribution from the outer shear layer and outer recirculation zone, as no flame is stabilized. It is interesting to note how the different locations drive and damp thermoacoustic instability between $\beta=90^o$ and $\beta=40^o$ for this high shear injector. Furthermore, as shown above, the jet tends to damp and drive thermoacoustic instability depending on the flow macrostructures.

\section{Conclusion}
This article scrutinized the flame macrostructures during the self-sustained thermoacoustic oscillation of a high shear swirl injector in a model can gas turbine combustor setup at atmospheric conditions. The flame macrostructures are altered using the flare angles of 90$^o$, 60$^o$, and 40$^o$. Depending on these flare angles, the relative location of recirculation zones, the axial extent of the jet and shear layers, and recirculation widths are modified. We explored the combustion dynamics of these flow features for fixed equivalence ratio (0.7) with inlet Reynolds numbers and thermal power ranging from 10506-16809 and 10.09-16.15 kW, respectively, as parameters. The combustion dynamics are characterized using acoustic pressure, high-speed stereo PIV, and high-speed OH$^*$ chemiluminescence measurements. The combustor exhibits dominant thermoacoustic instability in the frequency range of $140-210$ Hz, reaching a sound pressure level of around 140 dB.
Furthermore, we observed cycle-to-cycle variation of the acoustic pressure and heat release rate fluctuations caused by the flow turbulence. The flare angle of $\beta=40^o$ exhibits relatively large amplitude acoustic oscillations with compact flame and wider central recirculation zone. The phase averaged Rayleigh index highlights that the driving is mainly caused by the jets and inner shear layer for $\beta=40^o$. On the other hand, low amplitude acoustic oscillations with relatively slender flame and narrower central recirculation zones were observed for $\beta=90^o$. The phase averaged Rayleigh index shows that the inner recirculation zone and wake are the primary contributors for the driving of the thermoacoustic instability.  Furthermore, the flame stabilized in the jet tends to disrupt the thermoacoustic coupling, causing relatively more stable operation of the combustor for $\beta=90^o$ compared to the other two flare angles explored in this study.

\acknowledgement{Declaration of competing interest} \addvspace{10pt}
The authors declare no known competing interests.

%Use the acknowledgement environment defined in the template for this section, not \verb+\section*+.

\acknowledgement{Acknowledgments} \addvspace{10pt}
S.K.T and B.M. have contributed equally. The authors thank DRDO for financial support (Grant number 1-C-5).
% -------------------------------------------------------------------- %
% -------------------------------------------------------------------- %
% -------------------------------------------------------------------- %

 \footnotesize
 \baselineskip 9pt

% -------------------------------------------------------------------- %
% -------------------------------------------------------------------- %
% -------------------------------------------------------------------- %

\bibliographystyle{pci}
\bibliography{PCI_LaTeX}

% -------------------------------------------------------------------- %
% -------------------------------------------------------------------- %
% -------------------------------------------------------------------- %

\newpage

\small
\baselineskip 10pt

% -------------------------------------------------------------------- %
% -------------------------------------------------------------------- %
% -------------------------------------------------------------------- %

% -------------------------------------------------------------------- %
% -------------------------------------------------------------------- %
% -------------------------------------------------------------------- %

\end{document}